\documentclass{aa}
\usepackage{graphicx}
\usepackage{txfonts}
\newcommand{\TEFF}{\mbox{$T_{\rm eff}$}}
\newcommand{\LSOL}{\mbox{$L_{\sun}$}}
\newcommand{\micron}{\mbox{$\mu$m}}
\newcommand{\KMS}{\mbox{km s$^{-1}$}}
\newcommand{\HOH}{\mbox{H$_2$O}}
\newcommand{\cisotp}{\mbox{$^{12}$C/$^{13}$C}}
\begin{document}
\title{
Asymmetric silicate dust distribution 
toward \\
the silicate carbon star BM~Gem
\thanks{
Based on observations made with the Very Large Telescope Interferometer of 
the European Southern Observatory. Program ID: 078.D-0292}
}

\author{K.~Ohnaka\inst{1}
\and
H.~Izumiura\inst{2}
\and
Ch.~Leinert\inst{3}
\and
T.~Driebe\inst{1}
\and
G.~Weigelt\inst{1}
\and
M.~Wittkowski\inst{4}
}

\offprints{K.~Ohnaka}

\institute{
Max-Planck-Institut f\"{u}r Radioastronomie, 
Auf dem H\"{u}gel 69, 53121 Bonn, Germany\\
\email{kohnaka@mpifr-bonn.mpg.de}
\and
Okayama Astrophysical Observatory, National Astronomical Observatory, 
Kamogata, Asakuchi, Okayama 719-0232, Japan
\and
Max-Planck-Institut f\"ur Astronomie, 
K\"onigstuhl 17, 69117 Heidelberg, Germany
\and
European Southern Observatory, Karl-Schwarzschild-Str.~2, 
85748 Garching, Germany
}

\date{Received / Accepted }

\abstract
{
Silicate carbon stars show the 10~\micron\ silicate emission, 
despite their carbon-rich photospheres.  
They are considered to have circumbinary or circum-companion disks, 
which serve as a reservoir of oxygen-rich material shed by mass 
loss in the past.  
}
{We present $N$-band spectro-interferometric observations of the 
silicate carbon star BM~Gem using MIDI at the 
Very Large Telescope Interferometer (VLTI). 
Our aim is to probe the spatial distribution of oxygen-rich dust 
with high spatial resolution.  
}
{
BM~Gem was observed with VLTI/MIDI at 44--62~m baselines using the UT2-UT3 
and UT3-UT4 baseline configurations.  
}
{
The $N$-band visibilities observed for BM~Gem 
show a steep decrease from 8 to $\sim$10~\micron\ and a gradual increase 
longward of $\sim$10~\micron, reflecting the optically thin silicate emission 
feature emanating from sub-micron-sized amorphous silicate grains. 
The differential phases obtained at baselines of $\sim$44--46~m show 
significant non-zero values ($\sim \! 
-70$\degr) in the central part of the silicate emission feature between 
$\sim$9 and 11~\micron, revealing a photocenter shift and the asymmetric 
nature of the silicate emitting region.  
The observed $N$-band visibilities and differential phases 
can be fairly explained by a simple geometrical model in which the unresolved 
star is surrounded by a ring with azimuthal brightness modulation.  
The best-fit model is characterized by a broad ring ($\sim$70~mas across at 
10~\micron) with a bright region which is offset from the unresolved 
star by $\sim$20~mas at a position angle of $\sim$280\degr.  
This model can be interpreted as a system with a circum-companion disk and 
is consistent with the spectroscopic signatures of an accretion 
disk around an unseen companion recently discovered in the violet spectrum of 
BM~Gem. 
}
{}


\titlerunning{Asymmetric silicate dust distribution toward 
BM~Gem}
\authorrunning{Ohnaka et al.}
\maketitle

\section{Introduction}
\label{sect_intro}

``Silicate carbon stars'' are characterized by oxygen-rich circumstellar 
material as evident from the prominent silicate emission, 
despite their carbon-rich photospheres (Little-Marenin 
\cite{little-marenin86}; Willems \& de Jong \cite{willems86}).  
The detection of masers from O-bearing molecules such as OH and \HOH\ 
from some of the silicate carbon stars confirms 
the presence of oxygen-rich circumstellar gas (e.g., 
Nakada et al. \cite{nakada87}; Engels \cite{engels94}; 
Szymczak et al. \cite{szymczak01} and references therein).  
Moreover, optical spectroscopic studies (e.g., 
Lloyd-Evans \cite{lloyd-evans90}; Chan \cite{chan93}; 
Ohnaka \& Tsuji \cite{ohnaka99}) reveal that silicate carbon stars 
have \cisotp\ ratios as low as $\sim$4--5 (classified as ``J-type''), 
which is difficult to explain by the present stellar evolution 
theory.  
Currently, silicate carbon stars are considered to have an unseen, 
low-luminosity companion.  
The oxygen-rich material which was shed by mass loss when the primary star 
was an M giant can be stored in a circumbinary disk (Morris \cite{morris87}; 
Lloyd-Evans \cite{lloyd-evans90}) or in a circumstellar disk around the 
companion (``circum-companion disk'', Yamamura et al. \cite{yamamura00}), 
even after the primary star becomes a carbon star.  
In the latter case, Yamamura et al. (\cite{yamamura00}) suggest that 
the primary star drives an outflow from the circum-companion 
disk, and the silicate emission emanates from this outflow. 

Thanks to the recent progress of high-angular resolution observation 
techniques, 
there is growing observational evidence for the presence of 
circumbinary or circum-companion disks around silicate carbon stars.  
Our $N$-band spectro-interferometric observations of IRAS08002-3803 
with the MID-infrared Interferometric instrument (MIDI) at 
ESO's Very Large Telescope Interferometer (VLTI) 
spatially resolved the dusty environment of a silicate carbon star 
for the first time (Ohnaka et al. \cite{ohnaka06}), and 
our radiative transfer modeling suggests the presence of an optically 
thick circumbinary disk 
in which amorphous silicate and a second grain species (amorphous carbon, 
large silicate grains, or metallic iron) coexist.   
The presence of a circumbinary disk with multiple grain populations 
is also suggested for 
another silicate carbon star IRAS18006-3213 by Deroo et al. (\cite{deroo07}).  
The spatial distributions of the 22 GHz \HOH\ maser 
emission toward two silicate carbon stars, V778~Cyg and EU~And, have 
recently been resolved by Szczerba et al. (\cite{szczerba06}), Engels 
(priv. comm.), and 
Ohnaka \& Boboltz (\cite{ohnaka_boboltz08}) using MERLIN and the Very Long 
Baseline Array (VLBA).  
The observed spatio-kinematic structures are consistent with the presence of 
circum-companion disks. 
However, the formation mechanisms of these 
disks are little understood, and the possible link between the disk formation 
and the anomalously low \cisotp\ ratios is by no means clear, 
either.

Direct detection of a companion toward a silicate carbon star is hampered 
by the huge luminosity contrast between the primary carbon star 
($10^3$--$10^4$~\LSOL) and the companion ($\la$1~\LSOL), as well as small 
angular separations (e.g., Engels \& Leinert \cite{engels_leinert94}).  
However, Izumiura (\cite{izumiura03}) and Izumiura et al. (\cite{izumiura08}) 
have 
overcome the first problem by taking advantage of the fact that the violet 
flux of a carbon star (primary star) is heavily suppressed due to 
molecular and atomic absorption.  Using the High Dispersion Spectrograph (HDS) 
on the Subaru Telescope, they discovered continuum emission below 4000~\AA\ 
and Balmer lines showing P~Cygni profiles with an outflow velocity of 
$\ga$400~\KMS\ toward the silicate carbon star BM~Gem.  
These spectroscopic signatures strongly suggest the presence of an accretion 
disk around an unseen companion, 
although it could not be spatially resolved, and its nature---whether it is 
a main-sequence star or a white dwarf---remains to be determined.  
Comparison between the observationally derived accretion luminosity 
($\sim$0.2~\LSOL) and the theoretical prediction from the Bondi-Hoyle-type 
accretion resulted in estimated binary separations of $\sim$3--60~AU for a 
main-sequence star companion and $\sim$100--500~AU for a white dwarf 
companion.  
These values translate into angular separations of 2.5--50~mas and 
83--420~mas for a distance of 1.2~kpc adopted for BM~Gem by 
Izumiura et al. (\cite{izumiura08}).  

Given these estimated separations, it is challenging to resolve the 
circum-companion disk toward BM~Gem with a single-dish telescope.  
On the other hand, infrared interferometry provides us with a unique 
opportunity to achieve sufficiently high spatial resolution.  
In this paper, we present the results of mid-infrared spectro-interferometric 
observations of BM~Gem with VLTI/MIDI to probe 
the presence of the circum-companion disk with dust thermal emission.

\section{Observations}
\label{sect_obs}

\begin{table}
\begin{center}
\caption {MIDI observations of BM~Gem: 
night, time of observation (Coordinated Universal Time=UTC), telescope 
configuration (Tel.), projected baseline length $B_{\rm p}$, position angle 
of the projected baseline on the sky (P.A.), and seeing in the visible 
at the times of the 
observations of BM~Gem and the calibrators listed in Table~\ref{table_calib}. 
}
\begin{tabular}{r c c c r r l}
\hline
\hline
\# & Night & $t_{\rm obs}$ & Tel.      & $B_{\rm p}$ & P.A.  & Seeing \\ 
   &       & (UTC)         &            & (m)         & (\degr) & \\
\hline
1        & 2006 Dec. 29 & 07:29:47 & UT3-4 & 45.7 & 93.1  & 1\farcs0--1\farcs5\\
2        & 2006 Dec. 31 & 06:15:26 & UT3-4 & 55.2 & 99.1  & 0\farcs7--1\farcs0\\
3        & 2007 Jan. 04 & 06:31:59 & UT2-3 & 43.8 & 49.8  & 0\farcs8 \\
4        & 2007 Jan. 05 & 04:17:10 & UT3-4 & 62.1 & 108.0 & 0\farcs7--1\farcs0\\
\hline
\label{table_obs}
\end{tabular}
\end{center}
\end{table}

\begin{table}[!hbt]
\begin{center}
\caption {
MIDI observations of calibrators: 
12~\mbox{$\mu$m}\ fluxes ($F_{12}$), 
uniform-disk diameters ($d_{\rm{UD}}$), 
night, and the time stamp ($t_{\rm obs}$).  
The uniform-disk diameters were taken from the CalVin 
list available at ESO 
(http://www.eso.org/observing/etc/).  
The data set used for spectrophotometric calibration 
is marked with $\dagger$.  
}
\begin{tabular}{l r r c l}
\hline
\hline
Calibrator & $F_{12}$ & $d_{\rm{UD}}$  & night & $t_{\rm obs}$ (UTC) \\ 
           &  (Jy)    &  (mas)         &      &                     \\ \hline
HD48433    & 8.5      & $2.10\pm 0.13$ & 2006 Dec. 29 & 07:52:09 \\ 
           &          &                & 2006 Dec. 31 & 06:42:34$^{\dagger}$ \\ 
           &          &                & 2007 Jan. 04 & 06:08:26 \\ 
HD47205    & 9.7      & $2.30\pm 0.12$ & 2007 Jan. 05 & 03:41:13 \\
HD50778    & 24.6     & $3.95\pm 0.22$ & 2006 Dec. 31 & 01:35:21 \\ 
           &          &                & 2006 Dec. 31 & 05:35:01 \\
HD98430    & 19.1     & $3.09\pm 0.16$ & 2007 Jan. 04 & 07:03:45 \\ 

\hline

\label{table_calib}
\end{tabular}
\end{center}
\end{table}

Our MIDI observations of BM~Gem and the calibrators are summarized in 
Tables~\ref{table_obs} and \ref{table_calib}, respectively.  
BM~Gem was observed 
in the HIGH\_SENS mode using a prism ($\lambda/\Delta \lambda \simeq 30$ 
at 10~\micron), and 
four data sets were taken using the UT2-UT3 and UT3-UT4 
configurations with projected baseline lengths of 44--62~m.  
A detailed description of the instrument and the 
observing procedure are given in 
Przygodda et al. (\cite{przygodda03}), Leinert et al.\ (\cite{leinert04}), 
and Chesneau et al. (\cite{chesneau05}).  
We used the MIA+EWS package ver.1.5\footnote{Available at 
http://www.strw.leidenuniv.nl/\textasciitilde nevec/MIDI} 
to reduce the MIDI data (Leinert et al. \cite{leinert04}; 
Jaffe \cite{jaffe04}).  
While both of MIA and EWS can extract visibility amplitude, 
EWS can also extract the so-called 
``differential phase'', which contains information on the wavelength 
dependence of the object's phase.  However, it should be kept 
in mind that two pieces of information on the object's phase are 
lost in the derivation of differential phase as described in 
Jaffe (\cite{jaffe04}): the absolute phase offset and the phase gradient 
with respect to wavenumber.  

We assessed the data quality and derived the calibrated visibilities as 
described in Ohnaka et al. (\cite{ohnaka08}).  
Since it is difficult to properly estimate the errors of the calibrated 
visibilities from just 1--3 calibrator measurements, 
we assumed a total relative error of 15\% as 
adopted in Ohnaka et al. (\cite{ohnaka08}).  
The visibilities obtained with MIA and EWS are in good agreement, and 
we only show the results derived with EWS in the discussion 
below.  
We also extracted the absolutely calibrated $N$-band spectrum of BM~Gem 
from our MIDI data as described in Ohnaka et al. 
(\cite{ohnaka07}).  A calibrator observed at airmass similar to 
our target was used for the spectrophotometric calibration and is marked with 
$\dagger$ in Table~\ref{table_calib}.  
The error of the absolutely calibrated MIDI spectrum is difficult to estimate 
properly with only one calibrator, so we assumed errors of 
10--20\% as in Ohnaka et al. (\cite{ohnaka07}).

\section{Results}
\label{sect_result}

\begin{figure}[!hbt]
\resizebox{\hsize}{!}{\rotatebox{0}{\includegraphics{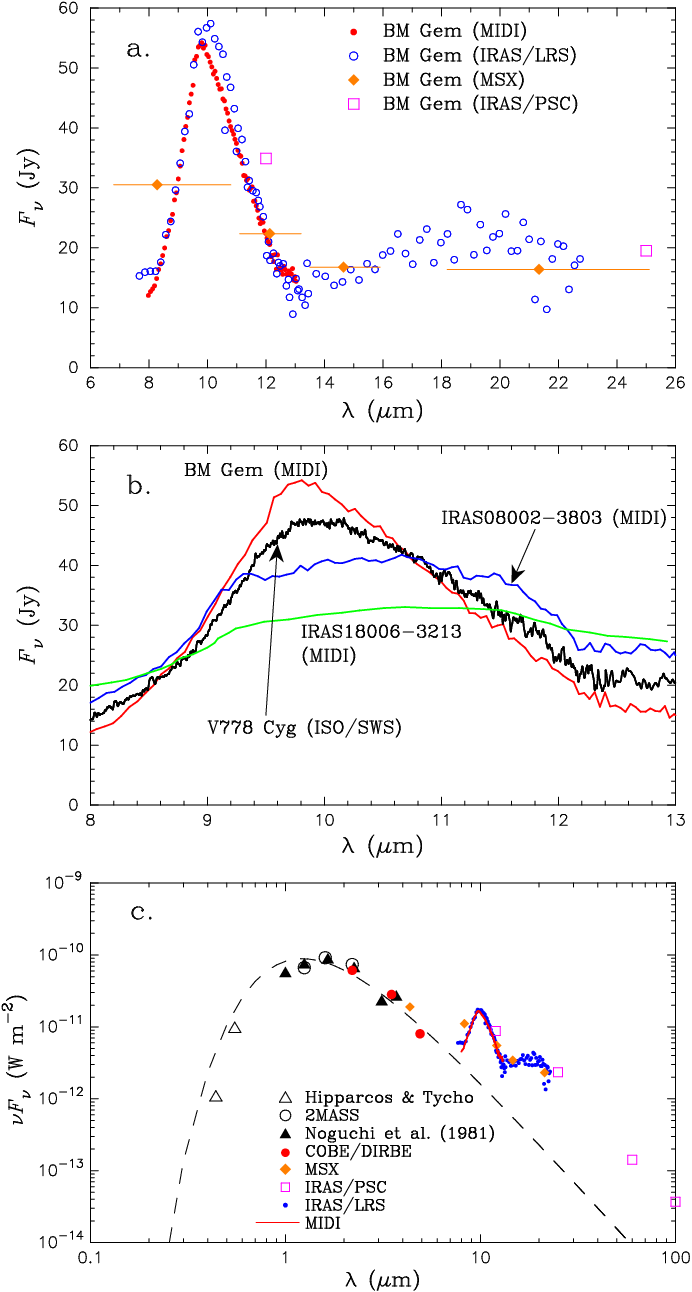}}}
\caption{
{\bf a:} $N$-band spectrum of BM~Gem extracted from the MIDI data, plotted 
together with IRAS/LRS as well as the photometric data from the MSX and 
IRAS PSC catalogs.  The width of each MSX band is shown with the horizontal 
bar.  
{\bf b:} Comparison of the $N$-band spectra of four silicate carbon stars: 
BM~Gem, V778~Cyg, IRAS08002-3803, and IRAS18006-3213.  
{\bf c:} SED of BM~Gem constructed from the (spectro)photometric data 
available in the literature (Hipparcos \& Tycho Catalogs, ESA \cite{esa97}; 
2MASS, Cutri et al. \cite{cutri03}; Noguchi et al. \cite{noguchi81}; 
COBE/DIRBE, Smith et al. \cite{smith04}; MSX; IRAS/PSC; IRAS/LRS; MIDI).  
The COBE/DIRBE data longward of 12~\micron\ are not used, because of the 
large errors.  
The dashed line represents the stellar flux contribution, which is 
approximated with a blackbody of 3000~K.  
}
\label{obsspec_bmgem}
\end{figure}

Figure~\ref{obsspec_bmgem}a shows the MIDI spectrum of 
BM~Gem, together with IRAS Low Resolution Spectrum 
(LRS)\footnote{Downloaded from 
http://staff.gemini.edu/\textasciitilde kvolk/getlrs\_plot.html
with the correction of the absolute calibration by 
Cohen et al. (\cite{cohen92}).} and the photometric data from the 
Midcourse Space Experiment (MSX, Egan et al. \cite{egan03}) and 
the IRAS Point Source Catalog (PSC).  
The MIDI spectrum and IRAS/LRS show that the flux level 
has little changed over the last 23 years, and the flux variation longward 
of 13~\micron\ is also absent, as can be seen from the IRAS/LRS and MSX data 
taken in 1996--1997.  
This stability of the mid-infrared flux is the same as found for V778~Cyg 
(Yamamura et al. \cite{yamamura00}), IRAS08002-3803 
(Ohnaka et al. \cite{ohnaka06}), IRAS18006-3213 (Deroo et al. \cite{deroo07}), 
and EU~And (Ohnaka \& Boboltz \cite{ohnaka_boboltz08}).  

In Fig.~\ref{obsspec_bmgem}b, we also plot the spectrum of V778~Cyg 
obtained with the Short Wavelength Spectrometer (SWS) onboard the 
Infrared Space Observatory (ISO) presented in 
Yamamura et al. (\cite{yamamura00}) as well as those of IRAS08002-3803 and 
IRAS18006-3213 obtained with MIDI by Ohnaka et al. (\cite{ohnaka06}) and 
Deroo et al. (\cite{deroo07}), respectively.  
BM~Gem and V778~Cyg show a well-defined peak at $\sim$9.8~\micron, while 
IRAS08002-3803 and IRAS18006-3213 show broader features.  
Yamamura et al. (\cite{yamamura00}) show that the silicate feature of V778~Cyg 
is characterized by optically thin emission from sub-micron-sized amorphous 
silicate grains.  The close resemblance between BM~Gem and V778~Cyg 
indicates the predominance of such small, amorphous silicate grains 
around BM~Gem, and the silicate dust toward this object is optically thin.  
This is also consistent with the spectral energy distribution 
(SED) of BM~Gem plotted in Fig.~\ref{obsspec_bmgem}c, which shows only a 
modest infrared excess longward of $\sim$8~\micron.  
As Yamamura et al. (\cite{yamamura00}) argue, such small grains in an optically 
thin environment are blown away by the radiation pressure of the primary 
(carbon-rich) star, suggesting that a circumbinary (or circum-primary) disk 
cannot exist for a long period of time.  This argument 
is in line with our interpretation of the $N$-band visibilities and 
differential phases presented in Sect~\ref{sect_discuss}.  

On the other hand, the broad $N$-band spectra of 
IRAS08002-3803 and IRAS18006-3213 indicate the presence of grain species 
other than small silicate.  
In particular, featureless dust emission from large grains, 
amorphous carbon, metallic iron, or iron sulfide is suggested for both 
objects (Ohnaka et al. \cite{ohnaka06}; Deroo et al. \cite{deroo07}).  
Also, the SEDs of these two silicate carbon stars show much larger infrared 
excesses longward of 3--4~\micron\ (see Fig.~5 in 
Ohnaka et al. \cite{ohnaka06} and Deroo et al. \cite{deroo07}) than BM~Gem. 
This suggests the presence of optically thick disks, which is actually 
confirmed by the radiative transfer modeling of 
Ohnaka et al. (\cite{ohnaka06}).  
These differences in mineralogy and optical thickness 
seem to affect the wavelength dependence of the observed $N$-band 
visibilities and differential phases, as discussed below. 

\begin{figure}[!hbt]
\resizebox{\hsize}{!}{\rotatebox{0}{\includegraphics{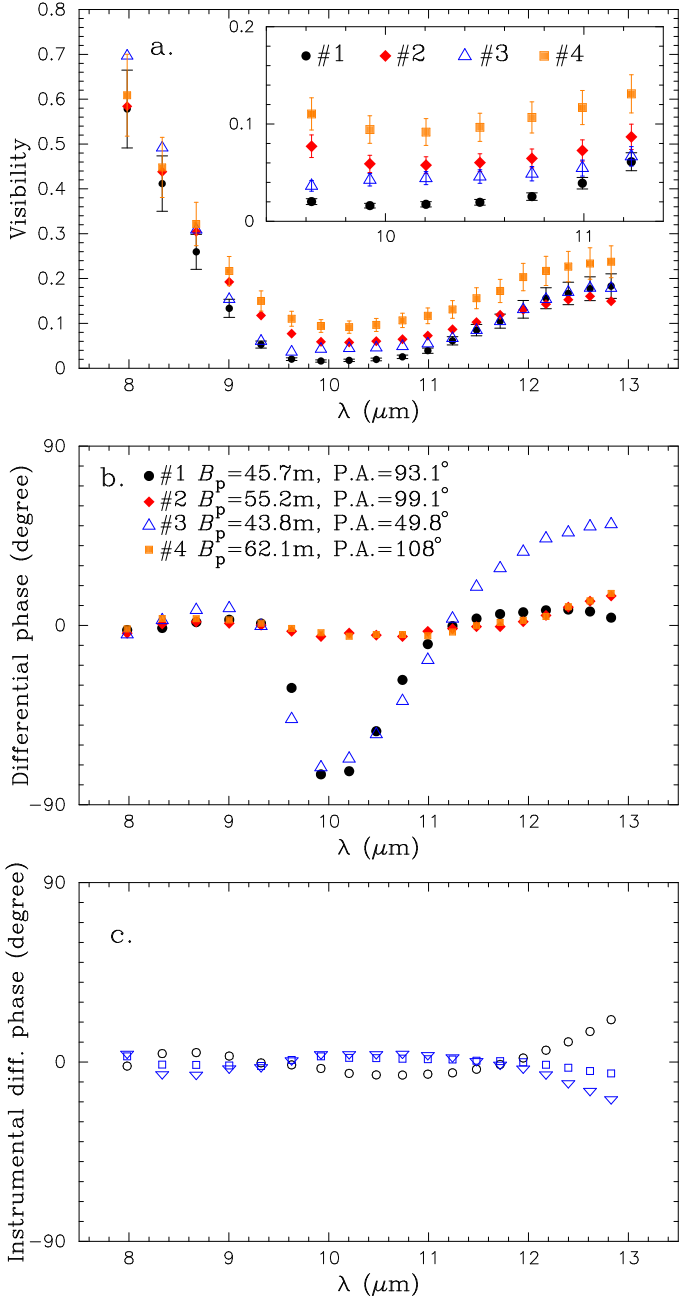}}}
\caption{
{\bf a:} $N$-band visibilities of BM~Gem obtained with MIDI.  
The error bars are 
shown only for the data sets \#1 and \#4 for visual clarity, but all 
error bars are drawn in the inset, which shows the enlarged view of the 
visibilities observed between 9.4 and 11.4~\micron.  The data set numbers 
are defined in Table~\ref{table_obs}.  
{\bf b:} Calibrated differential phases obtained for BM~Gem.  
{\bf c:} Instrumental differential phases as measured on the calibrators.  
The open circles represent the phase observed on the same night as the 
data set \#1, while the open squares and upside-down triangles represent 
that observed together with \#3.  
}
\label{obsvis_bmgem}
\end{figure}

The $N$-band visibilities observed toward BM~Gem shown 
in Fig.~\ref{obsvis_bmgem}a 
basically reflect the optically thin silicate emission: the visibilities are 
the lowest at $\sim$10~\micron, where the flux contribution of the extended 
silicate emission is the highest.  
The baseline dependence of the visibilities observed at $\la$9~\micron\ and 
at $\ga$11~\micron\ is not significant in the baseline range from 44 to 62~m, 
given the errors of the measurements.  
This means that the extended silicate emission is completely resolved out 
and that the sampled visibility component corresponds to the unresolved 
central star (its angular diameter is estimated to be $\sim$2~mas as explained 
below), for which the visibility is constant for the baselines used 
in our MIDI observations.  
In this case, the observed visibilities represent the fractional flux 
contribution from the star.  
On the other hand, the visibilities observed between 9 and 11~\micron\ 
increase with baseline length from 44-46~m to 62~m, 
as can be seen in 
the data \#1, \#2, and \#4 taken at roughly the same 
position angle.  
Therefore, the observed visibilities contain some information on 
the geometry of the silicate emitting region and will be modeled 
in the next section.

Figure~\ref{obsvis_bmgem}b shows the calibrated differential phases of 
BM~Gem.   
Significant non-zero differential phases ($\sim \! -70$\degr\ at 10~\micron) 
are detected in the central part of the silicate emission between 
$\sim$9 and 11~\micron\ in the data sets obtained at the shortest baselines 
(\#1 and \#3), while the differential phases observed at the longer baselines 
(\#2 and \#4) are close to zero.  
As shown in Fig.~\ref{obsvis_bmgem}c, the differential phases derived from 
the calibrators for the data sets \#1 and \#3 are close to zero.  
The phases of two calibrators for the data set \#3 
are in agreement within $\pm$5\degr, which we take as the error of 
the observed differential phases 
(the phase above 
12~\micron\ is more uncertain due to the increasing effect of the 
atmospheric water vapor).  
This confirms that the non-zero differential phases observed toward BM~Gem 
are real, revealing a photocenter shift in the silicate feature caused by 
an asymmetric circumstellar environment.  
The non-zero values are observed only 
in the central part of the silicate feature, although one might expect the 
differential phase to deviate from zero over the whole feature. 
However, as explained above, the silicate emission is completely 
resolved out at $\la$9~\micron\ and $\ga$11~\micron\ even at the shortest 
baselines.  The visibilities observed at these wavelengths correspond to 
the unresolved, thus symmetric star.  
This is why differential phases are nearly zero 
at these wavelengths, as well as in the data taken with the longer 
baselines.  

The wavelength dependence of the $N$-band visibilities observed for 
BM~Gem is in marked contrast to that previously observed toward 
IRAS08002-3803 and IRAS18006-3213 with MIDI by Ohnaka et al. (\cite{ohnaka06}) 
and Deroo et al. (\cite{deroo07}), respectively.  
The $N$-band visibilities of these two 
objects show an increase from 8 to $\sim$10~\micron\ and remain 
nearly constant or slightly decrease longward of 10~\micron.  
We also note that Deroo et al. (\cite{deroo07}) observed non-zero 
differential phases toward IRAS18006-3213, which are characterized by 
a significant phase jump at $\sim$8.3~\micron,  
completely different from that observed toward BM~Gem.  
Given that the visibility levels observed toward IRAS18006-3213 range from 
$\sim$0.1 to $\sim$0.7, the difference in the $N$-band visibilities and 
differential phases cannot be solely attributed to a difference in the 
object's angular size and/or baselines used in the observations.  
As outlined above, the dusty environment of BM~Gem is very different 
in mineralogy and optical thickness from that of IRAS08002-3803 and 
IRAS18006-3213.  
Therefore, the distinct $N$-band visibilities and differential phases 
observed toward BM~Gem most likely reflect the difference in dust chemistry 
and the structure of the circumstellar environment 
between BM~Gem and IRAS08002-3803/IRAS18006-3213.

\begin{figure*}[!hbt]
\resizebox{\hsize}{!}{\includegraphics[width=12cm]{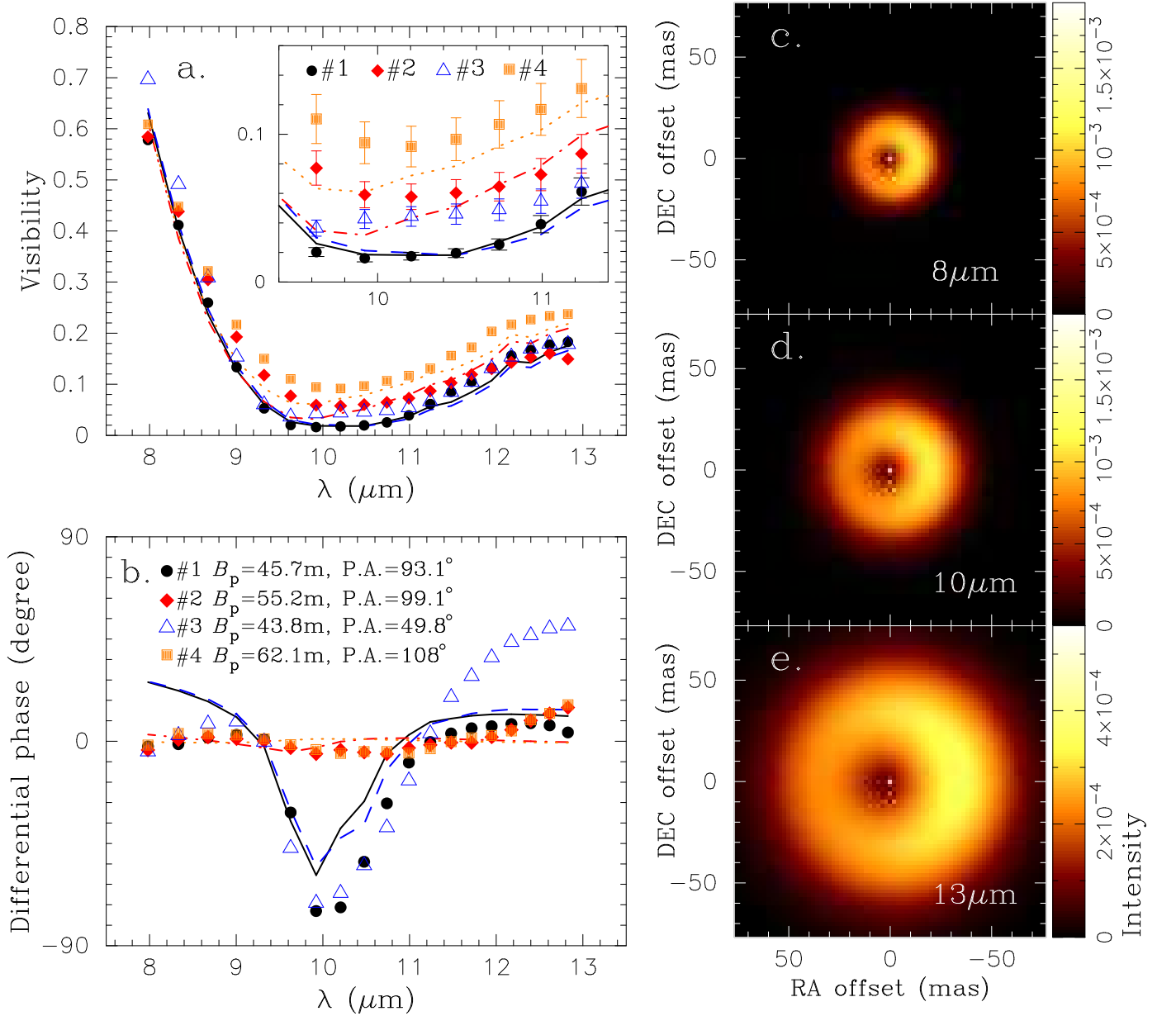}}
\caption{
{\bf a:} $N$-band visibilities of BM~Gem obtained with MIDI (symbols) 
and those predicted by the best-fit asymmetric-disk model discussed in 
Sect.~\ref{sect_model} (lines).  
The error bars are shown only in the inset, which shows the enlarged view of 
the visibilities observed between 9.4 and 11.4~\micron.  The data set numbers 
are defined in Table~\ref{table_obs}.  
The model visibilities are shown by the solid, dashed-dotted, dashed, and 
dotted lines for the data sets \#1, \#2, \#3, and \#4, respectively, 
with the same colors as the corresponding data sets.  
{\bf b:} Calibrated differential phases obtained for BM~Gem (symbols) and 
those predicted by the asymmetric-disk model (lines) shown in the same 
manner as in the panel {\bf a}.  
The dashed-dotted and dotted lines (red and orange lines) are almost entirely 
overlapping. 
{\bf c--e:} 8, 10, and 13~\micron\ images predicted by the best-fit model.  
The radius of the ring is 20~mas with a Gaussian radial cross 
section with a FWHM of 27~mas at 10~\micron.  
The azimuthal intensity modulation is 
characterized by $a$ = 0.2 and $\Phi$ = 280\degr.  
The image at each wavelength is normalized so that the integrated flux is 
1 (the color on the central star is saturated).  
A color version of this figure is available in the electronic 
edition.  
}
\label{model_bmgem}
\end{figure*}

\section{Geometrical model}
\label{sect_model}

To characterize the asymmetric silicate emitting region, 
we adopted an ``asymmetric-disk'' model, in which 
an unresolved star is surrounded by a ring with a Gaussian radial cross 
section, and the azimuthal brightness distribution of the ring is sinusoidally 
modulated.  
This model was used to interpret the interferometric data on 
Herbig Ae/Be stars by Monnier et al. (\cite{monnier06}).  
The intensity at the radial distance $r$ from the star and the position angle 
$\phi$ is expressed as 
\[
I_{\lambda}(r, \phi) \propto e^{-\left(\frac{r-R_{\lambda}}{\sigma_{\lambda}}\right)^2}
\times \{ a \cos (\phi - \Phi) + 1\}, 
\]
where $R_{\lambda}$ and $\sigma_{\lambda}$ denote the 
radius and the Gaussian width of the ring, respectively.  
The intensity contrast between 
the brightest and faintest points of the ring is determined by the parameter 
$a$, and $\Phi$ is the position angle of the brightest point.  
$R_{\lambda}$ and $\sigma_{\lambda}$ are dependent on wavelength, 
but they cannot be well constrained by the current MIDI data at $\la$9~\micron\ 
and $\ga$11~\micron, because the extended component is resolved out at these 
wavelengths.  To take into account that the emission at the longer 
wavelengths originates in the cooler, outer region, we assumed the radius to be 
expressed as 
\[
R_{\lambda} = R_{10\mu{\rm m}} \times \left(\frac{\lambda}{10}\right)^p, 
\]
where $R_{10\mu{\rm m}}$ denotes the ring radius at 10~\micron.  
The ratio between $R_{\lambda}$ and $\sigma_{\lambda}$ was assumed to be 
independent of wavelength.  
Therefore, the free parameters of the model are $R_{10\mu{\rm m}}$, $p$, 
the ratio $\sigma_{\lambda}/R_{\lambda}$, $a$ and $\Phi$.  

We first derived the ratio of the stellar flux to the total flux at each 
wavelength as follows.  If we approximate 
the stellar radiation with a blackbody of effective temperature \TEFF\ 
(3000~K for BM~Gem, Ohnaka \& Tsuji \cite{ohnaka99}), 
the observed flux $F_{\lambda}^{\rm obs}$ (star + extended dust emission) is 
given by 
\[
F_{\lambda}^{\rm obs} = \theta_{\star}^2 B_{\lambda}(\TEFF)/f_{\lambda}^{\star}, 
\]
where $\theta_{\star}$ and $f_{\lambda}^{\star}$ are the stellar angular radius 
and the ratio of the stellar flux to the total flux, respectively.  
As mentioned above, the visibilities at $\la$9~\micron\ and $\ga$11~\micron\ 
represent the fractional flux contribution of the star.  
Using the observed 8~\micron\ visibility for $f_{\lambda}^{\star}$ 
and the observed 8~\micron\ flux, $\theta_{\star}$ was 
derived to be 1.1~mas.  Using this $\theta_{\star}$ and the MIDI 
spectrum, we determined $f_{\lambda}^{\star}$ for all wavelengths in the 
$N$ band.  
We note here that as Fig.~\ref{obsspec_bmgem}c shows, the SED of BM~Gem 
shortward of $\sim$5~\micron\ is well fitted by the blackbody of 3000~K, 
suggesting the absence of hot dust, which would have manifested itself as an 
additional (unresolved) component in our MIDI observations.  
This verifies our assumption that the observed visibility at 8~\micron\ 
represents the fractional flux of the star.  

The intensities of the point-source-like star and the extended 
component at each wavelength were set so that the flux ratio 
is consistent with the derived $f_{\lambda}^{\star}$.  
The visibility amplitude and phase were calculated from 
the two-dimensional complex visibility computed at each wavelength.  
The differential phase was calculated in the same manner as 
the EWS software derives it from MIDI data: 
a least-square-fit with a straight line (i.e., a linear function with 
respect to wavenumber) was performed on the model phase as a function 
of wavenumber.  The differential phase was then derived by subtracting the 
fitted linear component from the model phase.  
The sign of differential phase measured with MIDI is defined as described 
in Ratzka et al. (\cite{ratzka08}).  In the model fitting, we excluded 
the observed differential phase data above 12~\micron\ because of the 
larger systematic uncertainties.

Figures~\ref{model_bmgem}a and \ref{model_bmgem}b show a comparison between 
the best-fit model and the MIDI data.  
This model is characterized by a broad ring with a radius of 20~mas 
and a Gaussian radial cross section with a FWHM of 27~mas at 10~\micron\ 
($\sigma_{\lambda}/R_{\lambda}$ = 0.8).  The wavelength dependence of the ring 
radius is characterized by $p = 2$, which results in overall sizes of 
$\sim$45, 70, and 120~mas at 8, 10, and 13~\micron, respectively, 
as shown in Figs.~\ref{model_bmgem}c--\ref{model_bmgem}e.  
The intensity of the ring has a mild azimuthal modulation with $a$ = 0.2, 
and the brightest point is offset by 20~mas (at 10~\micron) at a position 
angle of 280\degr\ with respect to the star.   
Figure~\ref{model_bmgem}a reveals that the observed $N$-band visibilities 
are reasonably reproduced by the model.  
As shown in Fig.~\ref{model_bmgem}b, the predicted differential phases also 
agree fairly with the observed data.  
The model predicts the non-zero differential phase values to be 
somewhat too large at $\la$9~\micron\ and too small at $\ga$9~\micron\ 
compared to the observed data.  Also, the model visibility for the data set 
\#3 is lower than observed.  
However, given the simple nature of the model, the fit can be regarded as 
fair.  
The $\chi^2$ value is 3.8 for this best-fit model.  
The formal errors of the ring radius at 10~\micron\ ($R_{10\mu {\rm m}}$), 
$p$, the ratio $\sigma_{\lambda}/R_{\lambda}$, azimuthal brightness 
modulation ($a$), and the position angle of the brightest point ($\Phi$) are 
$\pm 2$~mas, $\pm 0.5$, $\pm 0.05$, $\pm 0.1$, and $\pm 40$\degr, respectively.  
However, it should be kept in mind that 
the extended silicate emission is nearly resolved out by our 
MIDI observations, which makes it difficult to glean information on 
the overall geometry of the silicate emitting region.  
Therefore, the above model parameters would better be regarded as 
preliminary despite the rather small formal errors, 
until they are confirmed by future MIDI observations with 
a better $(u,v)$ coverage.

\section{Discussion}
\label{sect_discuss}

Given the spectroscopic signatures of a circum-companion accretion disk 
discovered toward BM~Gem by Izumiura (\cite{izumiura03}) and 
Izumiura et al. (\cite{izumiura08}), it is tempting to interpret the above 
model as a circum-companion disk.  
In the scenario of Yamamura et al. (\cite{yamamura00}), 
the radiation pressure from the primary star drives an outflow from the 
circum-companion disk.  
Depending on the orbital period of the binary system and the outflow velocity, 
the material blown away from the disk may trail like a cometary tail and show a 
spiral structure surrounding the whole binary system.   
The material surrounding the whole system (corresponding to the 
``cometary tail'') can be represented by the broad ring of the asymmetric-disk 
model, while the circum-companion disk (the ``coma'' of the cometary 
trail) appears as the offset bright region.  
In our model, the offset of the bright feature is dependent on 
wavelength as the ring's radius, although the position of the companion 
is independent of wavelength.  However, since the mid-infrared emission 
reflects the distribution of dust, not exactly the companion itself, such 
wavelength dependence of the offset may also be expected.

If the above model represents the circum-companion disk, 
the offset of the bright feature from the central star, 13--34~mas, 
corresponds to the angular separation between the primary star 
and the companion.  
This favors a main-sequence star companion rather than a white dwarf 
companion, because the luminosity from an accretion disk around a white 
dwarf at separations of 13--34~mas (16--41~AU at a distance of 1.2~kpc) 
is estimated to be $\sim$5--18~\LSOL\ in 
the Bondi-Hoyle formulation (Eq.~5 in Izumiura et al. \cite{izumiura08}). 
This is significantly higher than the observationally derived value of 
$\sim$0.2~\LSOL\ (0.03--0.6~\LSOL, Izumiura et al. \cite{izumiura08}).  
However, it should be noted that the Bondi-Hoyle-type accretion could 
overestimate the accretion rate as much as by an order of magnitude 
(e.g., Nagae \cite{nagae04} and references therein).  
In this case, the predicted accretion luminosity is $\sim$0.5--1.8~\LSOL, 
which is marginally compatible with the observed range.  
Therefore, a main-sequence star companion seems more likely, although 
a white dwarf companion cannot be entirely excluded.

\section{Conclusion}
\label{sect_concl}

Our VLTI/MIDI observations of the silicate carbon star BM~Gem have revealed 
that the $N$-band visibilities are characterized by a steep 
decrease from 8 to $\sim$10~\micron\ and a gradual increase longward of 
10~\micron, reflecting the optically thin 10~\micron\ emission feature 
emanating from sub-micron-sized amorphous silicate grains.  
Furthermore, we have detected significant non-zero differential phases 
($\sim \! -70$\degr) between $\sim$9 and 11~\micron, which reveals the 
asymmetric nature of the silicate emitting region.  
The wavelength dependence of the observed $N$-band visibilities and 
differential phases is remarkably different from that previously observed 
toward two silicate carbon stars, IRAS08002-3803 and IRAS18006-3213.  
This is most likely to reflect the differences in mineralogy, optical 
thickness, and geometry (circumbinary or circum-companion) of the disks.

Our simple geometrical modeling of the observed visibilities and differential 
phases suggests the presence of a bright region which is offset from the 
central star by 13--34~mas (16--41~AU at 1.2~kpc) at a position angle of 
$280 \pm 40$\degr.  
This bright region can be interpreted as a circum-companion disk.  
The derived offset from the central star makes a white dwarf companion 
unlikely (thus leaving a main-sequence star companion more favorable), 
because a white dwarf companion at these separations would produce 
too much accretion luminosity compared to that observed.  

However, the lack of detailed information on the geometry of the 
silicate dust distribution makes this circum-companion 
disk interpretation not yet definitive.  For example, we cannot yet 
entirely exclude the possibility that the asymmetric-disk model 
represents a circumbinary disk with an asymmetric structure. 
MIDI observations with shorter baselines are indispensable for putting tighter 
constraints on the size and shape of the silicate emitting region. 
Also, high-resolution imaging at violet/blue wavelengths 
is important for direct detection of the 
circum-companion accretion disk.  Combination of the violet/blue 
high-resolution imaging and MIDI observations will provide definitive 
confirmation that silicate dust is stored in a circum-companion disk.

\begin{acknowledgement}
We thank the ESO VLTI team on Paranal and in Garching and the MIDI 
team for carrying out the observations and making the data reduction 
software publicly available.  
We also thank the referee, H.~van~Winckel, for his constructive comments.

\end{acknowledgement}


\begin{thebibliography}{}


\bibitem[1993]{chan93}
Chan, S. J. 1993, AJ, 106, 2126

\bibitem[2005]{chesneau05}
Chesneau, O., Meilland, A., Rivinius, T., et al.\ 2005, A\&A, 435, 275

\bibitem[1992]{cohen92}
Cohen, M., Walker, R. G., \& Witteborn, F. C.\ 1992, AJ, 104, 2030

\bibitem[2003]{cutri03}
Cutri, R. M., Skrutskie, M. F., Van Dyk, S., et al. 2003, 
The 2MASS All-Sky Catalog of Point Sources

\bibitem[2007]{deroo07}
Deroo, P., Van Winckel, H., Verhoelst, T., et al. 2007, A\&A, 467, 1093

\bibitem[1997]{esa97}
ESA 1997, The Hipparcos and Tycho Catalogues (ESA SP-1200)

\bibitem[2003]{egan03}
Egan, M. P., Price, S. D., Kraemer, K. E., et al. 2003, 
The Midcourse Space Experiment Point Source Catalog Version 2.3, 
Air Force Research Laboratory Technical Report AFRL-VS-TR-2003-1589

\bibitem[1994]{engels94}
Engels, D. 1994, A\&A, 285, 497

\bibitem[1994]{engels_leinert94}
Engels, D., \& Leinert, Ch. 1994, A\&A, 282, 858

\bibitem[2003]{izumiura03}
Izumiura, H. 2003, Ap\&SS, 283, 189

\bibitem[2008]{izumiura08}
Izumiura, H., Noguchi, K., Aoki, W., et al. 2008, ApJ, in press, 
astro-ph/0804.4040

\bibitem[2004]{jaffe04}
Jaffe, W.\ 2004, SPIE Proc., 5491, 715

\bibitem[2004]{leinert04}
Leinert, Ch., van Boekel, R., Waters, L. B. F. M., et al.\ 2004, 
A\&A, 423, 537 

\bibitem[1986]{little-marenin86}
Little-Marenin, I. R.\ 1986, ApJ, 307, L15

\bibitem[1990]{lloyd-evans90}
Lloyd-Evans, T.\ 1990, MNRAS, 243, 336

\bibitem[2006]{monnier06}
Monnier, J. D., Berger, J.-P., Millan-Gabet, R., et al. 2006, ApJ, 647, 444

\bibitem[1987]{morris87}
Morris, M.\ 1987, PASP, 99, 1115

\bibitem[2004]{nagae04}
Nagae, T., Oka, K., Matsuda, T., et al.\ 2004, A\&A, 419, 335

\bibitem[1987]{nakada87}
Nakada, Y., Izumiura, H., Onaka, T., et al.\ 1987, ApJ, 323, L77

\bibitem[1981]{noguchi81}
Noguchi, K., Kawara, K., Kobayashi, Y., et al. 1981, PASJ, 33, 373

\bibitem[2008]{ohnaka_boboltz08}
Ohnaka, K., \& Boboltz, D. A. 2008, A\&A, 478, 809

\bibitem[1999]{ohnaka99}
Ohnaka, K., \& Tsuji, T.\ 1999, A\&A, 345, 233

\bibitem[2006]{ohnaka06}
Ohnaka, K., Driebe, T., Hofmann, K.-H., et al. 2006, A\&A, 445, 1015

\bibitem[2007]{ohnaka07}
Ohnaka, K., Driebe, T., Weigelt, G., \& Wittkowski, M. 2007, A\&A, 466, 1099

\bibitem[2008]{ohnaka08}
Ohnaka, K., Driebe, T., Hofmann, K.-H., Weigelt, G., \& Wittkowski, M.\ 
2008, A\&A, 484, 371

\bibitem[2003]{przygodda03}
Przygodda, F., Chesneau, O., Graser, U., Leinert, Ch., \& 
Morel, S.\ 2003, Ap\&SS, 286, 85

\bibitem[2008]{ratzka08}
Ratzka, T., et al. 2008, in preparation

\bibitem[2004]{smith04}
Smith, B. J., Price, S. D., \& Baker, R. I.\ 2004, ApJS, 154, 673

\bibitem[2006]{szczerba06}
Szczerba, R., Szymczak, M., Babkovskaia, N., et al. 2006, A\&A, 452, 561

\bibitem[2001]{szymczak01}
Szymczak, M., Szczerba, R., \& Chen, P. S. 2001, In: Post-AGB Objects as a 
Phase of Stellar Evolution, eds. R.~Szczerba and S.~K.~G\'{o}rny, 
Ap\&SS Library Vol. 265, p.439

\bibitem[2000]{yamamura00}
Yamamura, I., Dominik, C., de Jong, T., Waters, L. B. F. M., \& 
Molster, F. J.\  2000, A\&A, 363, 629

\bibitem[1986]{willems86}
Willems, F., \& de Jong, T.\ 1986, ApJ, 309, L39


\end{thebibliography}
\end{document}